\documentclass[aps,prc,twocolumn,tightenlines,superscriptaddress,nofootinbib,a4paper]{revtex4-2}
\usepackage{graphicx,color}
\usepackage{braket}
\usepackage{amsmath}
\usepackage{amssymb,amsbsy}
\usepackage[T1]{fontenc}

\newcommand{\be}{\begin{eqnarray}}
\newcommand{\ee}{\end{eqnarray}}
\newcommand{\ben}{\begin{eqnarray*}}
\newcommand{\een}{\end{eqnarray*}}

\newcommand{\prj}{{}^{29}{\rm Ne}}
\newlength{\figwidth}
\def\vec#1{\boldsymbol{#1}}

\begin{document}

\title{Single-nucleon removal cross sections on nucleon and nuclear targets}

\author{J.\,A. Tostevin}
\affiliation{Department of Physics, Tokyo Institute of Technology,
2-12-1 O-Okayama, Meguro, Tokyo 152-8551, Japan}
\affiliation{Department of Physics, Faculty of Engineering and Physical
Sciences, University of Surrey, Guildford, Surrey GU2 7XH, United Kingdom}
\date{\today}

\begin{abstract}
The eikonal direct-reaction model, as used in spectroscopic studies of
intermediate-energy nucleon-removal reactions on light target nuclei, is
considered in the case of a proton target and applied to neutron removal
from $^{29}$Ne at 240 MeV/nucleon. The computed cross sections and their
sensitivities are compared using an earlier detailed analysis of carbon
target data. The nuclear structure input, for the $^{29}$Ne ground-state
and $^{28}$Ne final states, is that deduced from the carbon target analysis.
The comparisons quantify the sensitivity of the two reactions to the angular
momenta and binding energies of the active valence orbitals - showing the
carbon target to be relatively more efficient for removals from weakly-bound,
low-$\ell$, halo-like orbitals. Probing this sensitivity experimentally would
provide useful tests of these predictions and of the model's description
of the reaction mechanism.
\end{abstract}
\maketitle

\section{Introduction}
The evolving shell structure and resulting spectroscopy seen in very neutron-rich
nuclei in the region of the Island of Inversion \cite{wbb} near $Z=10$ is
complex. Extensive studies in this demanding region of the chart of nuclides,
see e.g. \cite{GG}, have provided challenging tests of large-scale shell-model
calculations and have stimulated and guided the development of improved
shell-model effective interactions, e.g. Refs.\ \cite{Uts01,Pov12,Cau14,ekk}.
Data sets and analyses of numerous intermediate-energy ($80-250$ MeV/nucleon)
nucleon-removal reactions on both lead and light nuclear targets, such as carbon and
beryllium, have proved highly effective in: (a) identifying the active valence
single-particle orbitals near the neutron and proton Fermi surfaces, and (b)
mapping their migration with $A$ and $Z$ in such nuclei far from stability.
Developments in reaction targets and experimental capabilities, e.g. \cite{Minos},
have generated an increased interest in such fast, direct nucleon-removal
(or knockout) reactions induced on a proton target.
In this work, our focus is on direct-reaction model calculations for fast
nucleon removal reactions from the projectile $^{29}$Ne on a proton target
at a reaction energy of 240 MeV/nucleon. The $^{29}$Ne system is chosen:
(a) as detailed comparisons are possible with an earlier analysis of removal
reaction data on a carbon (and lead) target \cite{Kob29} performed at the same
beam energy, and (b) $^{29}$Ne is rather typical if weakly-bound, neutron-rich
systems in this mass/charge region, with occupancies of both $sd$-shell and
$pf$-shell (intruder) valence orbitals.

As in the model treatment of reactions on light nuclear targets, usually
carbon or beryllium, the basis of calculations is the use of: (i) the eikonal
(forward scattering) and sudden (fast collision) approximations to the collision
dynamics, combined with (ii) spectroscopic factors from shell-model wave function
overlaps, which determine the strengths of the active configurations of the
removed nucleons \cite{Tos01,HaT03}. Importantly, the use of a proton target
fundamentally alters the dominant direct reaction mechanism and thus the
reaction inputs, parameter choices, and also, potentially, the resulting
structural and parameter sensitivities.

\section{Direct reaction model}
In the eikonal-model description of inverse-kinematics single-nucleon
removal reactions from a mass $A$ projectile incident on a nuclear target,
removal events result from both elastic and inelastic interactions of the
struck (removed) nucleon with the target. As the target final-states from
these two removal mechanisms are distinct, if only the mass $A$$-$1
residual nucleus is detected, these two (incoherent) contributions to the
single-particle removal cross section must be summed, i.e. $\sigma_\text{sp}=
\sigma_\text{inel}+\sigma_\text{elas}$. In this sudden, eikonal-model
approach, details of which have been presented elsewhere \cite{Tos01},
the S-matrices that enter the approximate expressions for $\sigma_\text{inel}$
and $\sigma_\text{elas}$ describe the effects of the complex
optical-model interactions between the removed valence nucleon and the
mass $A$$-$1 reaction residue with the target. These S-matrices account
for the loss of flux of the fast, forward-traveling nucleon and residual
nucleus due to scattering and absorption by these optical potentials.
For reactions on a target of light composite nuclei, the calculated
inelastic (also called stripping) mechanism is the more important, and
$\sigma_\text{inel}$ dominates the removal cross section. These fractions
of inelastic and elastic removal events, sensitive to the nucleon's
separation energy, have also been measured in several cases and shown to
be very well described by the model calculations for reactions involving
nucleon removal from both well-bound ($\approx$17 MeV) and weakly-bound
($\lesssim$1 MeV) valence orbitals \cite{Bazin,Wimmer}.

By contrast, in the case of a proton target: (i) the removed nucleon-target
S-matrix describes the nucleon-nucleon (NN) system, and (ii) at the energies
of interest here the proton is inert. The relative importance of the two
reaction mechanisms is thus completely altered and the removal cross section
is now determined by $\sigma_\text{elas}$. Thus, the use of a proton
target also permits analyses based on a number of alternative direct
reaction model approaches that are suited to breakup studies - including the
continuum-discretized coupled-channels (CDCC) methodology \cite{cdcc}
and more recently developed coupled-channels approaches \cite{Ramos} to
the $(p,2p)$ and $(p,pn)$ processes. The CDCC approach was used successfully
by Kondo {\em et al.} \cite{Kondo} in the case of neutron removal reactions
from $^{18,19}$C on a proton target. Quasi-free, distorted-waves impulse
approximation (DWIA) nucleon knockout models have also been explored and
applied with some success for a limited range of projectiles--primarily
the oxygen and carbon isotopes--see for example Refs.\ \cite{qe1,qe2,qe3}.
Here, the effects of the change in the dominant reaction mechanism are
studied using the more computationally efficient eikonal-model approach
-- that derives a closed-form expression \cite{Tos01} for the elastic breakup
cross section of interest, integrated over all continuum final
states of the residue and removed nucleon. The residual nucleus momentum
distributions from this model, on a proton target, were used successfully
to identify the angular momenta of $^{28}$F final states in Ref. \cite{Revel},
but no comparisons of cross sections were available there. A similar model
to that used here was applied to the $^{17}$C$(-n)$ reaction at the lower
beam energy of 70 MeV/nucleon \cite{Satou}, but these calculations were not
benchmarked against nuclear target data, the primary interest here. We
compare the sensitivity of the modelled cross sections on a proton and nuclear
target and quantify their differences, both as potential probes of the reaction
mechanisms and for use in spectroscopy.

\subsection{Nucleon-nucleon description}
The required NN S-matrix, describing the removed nucleon-target proton
collisions, will be denoted $S_{jp}(b)$ where $j$ denotes the species of
the removed nucleon, i.e. $j=$ n,p. This NN scattering operator, expressed
as a function of the impact parameter, $b$, of the NN relative motion,
is conventionally written \cite{AT}
\begin{eqnarray}
S_{jp}(b) =1-\Gamma_{jp} (b)
\end{eqnarray}
where $\Gamma_{jp}$, the NN profile function, is determined by the two-dimensional
(2D) Fourier transform of the NN scattering amplitude $f_{jp}({q})$. That is
\begin{eqnarray}
\Gamma_{jp}(b)=\frac{1}{2 i\pi k}\int d^2\vec{q}\,e^{-i\vec{q}\cdot \vec{b}}\,
f_{jp}({q})\,,\label{transf}
\end{eqnarray}
where $\vec{q}$, the momentum transfer, is in the plane perpendicular to the
beam direction. As is often used, the profile function is parameterized as
\begin{eqnarray}
\Gamma_{jp}(b)=\frac{\sigma_{jp}}{2 i}\left(i+\alpha_{jp}\right) g_2
(\beta_{jp},b)\,
\end{eqnarray}
where $g_2(\beta,b)$ is the normalized two-dimensional (2D) Gaussian form factor
\begin{eqnarray}
g_2(\beta,b)=\frac{1}{2 \pi\beta}\exp(-b^2/2\beta)
\end{eqnarray}
representing the finite-range of the NN interaction. Thus, from the inverse
transform of Eq. (\ref{transf})
\begin{eqnarray}
f_{jp}(q)=\frac{k}{4\pi}\sigma_{jp}\left(i+\alpha_{jp}\right) \exp(-
\beta_{jp}q^2/2)\,\label{amp}
\end{eqnarray}
and, from the optical theorem identity, namely
\begin{eqnarray}
{\rm Im.}\,f (0^\circ)={\rm Im.}\,f (q=0)=\frac{k}{4\pi}\sigma_{tot}\,,
\label{optical}
\end{eqnarray}
one notes that the $\sigma_{jp}$ are the np and pp total cross sections. These
are computed here from the Charagi and Gupta parameterization \cite{CG} of
the experimental NN data, giving $\sigma_{pp}=22.02$ mb and $\sigma_{np}
=37.14$ mb at the relevant laboratory energy of 240 MeV per nucleon.
Also clear from Eq. (\ref{amp}), the parameters $\alpha_{jp}$ are the ratios
of the real to the imaginary parts of the NN forward-scattering amplitudes,
$f_{jp}(q=0)$, while $\beta_{jp}$ determines the range of the assumed Gaussian
form factor (of Gaussian range $\gamma_{jp} =\sqrt{2\beta_{jp}}$). Here, the
$\alpha_{jp}$ are interpolated from the tabulation (on the interval 100--1000
MeV) of Ray \cite{Ray}, giving values $\alpha_{np}=0.533$ and $\alpha_{pp}=
0.991$. For the range parameters $\beta_{jp}$, we follow Ref. \cite{abu} and
dictate, since the NN interaction energy is below the pion production threshold
and the NN scattering is entirely elastic, that the total and total elastic
cross sections derived from the NN S-matrices are equal. This requirement
fixes the range parameter $\beta_{jp}$ such that
\begin{eqnarray}
\beta_{jp}=\frac{\sigma_{jp}\left(1+\alpha_{jp}^2\right)}{16\pi}\,.
\label{betaeq}
\end{eqnarray}
In section \ref{sens} we assess the sensitivity of our calculated removal
reaction cross-sections to this choice of $\alpha_{jp}$ (and hence $\beta_{jp}$).
It will be shown there that this NN parameter sensitivity is actually very weak.

\subsection{Target proton-residue interactions}
The remaining dynamical input to the removal cross section is the eikonal
S-matrix that describes the interaction of the mass $A$$-$1 reaction residue
with the proton target. This is computed in the optical limit, or $t \rho$
folding approximation to the proton-residue optical potential. This potential,
and its S-matrix, includes the effects of the size and neutron-proton asymmetry
of the residual nucleus ($r$) through its point-neutron and proton one-body
densities $\rho^{(j)}_r$. The potential used is thus
\begin{eqnarray}
{\cal U}_{pr}(R)= \sum_{j=n,p}\,\int d \vec{r} \rho^{(j)}_r(r)
t_{jp}(\vert\vec{R}+\vec{r}\vert)~,
\end{eqnarray}
where the NN effective interaction, $t_{jp}$, consistent with our treatment
of the NN profile function and S-matrix given above, is
\begin{eqnarray}
t_{jp}(r)=-\frac{\hbar v}{2}\sigma_{jp} \left(i+\alpha_{jp}\right) g_3
(\beta_{jp},r)\,.
\end{eqnarray}
Here, $v$ is the residue-proton relative velocity, $g_3(\beta,r)$
is a normalized 3D Gaussian function with range parameter $\beta$, and the
parameters $\sigma_{jp}$, $\alpha_{jp}$ and $\beta_{jp}$ are the same as
were stated above.

The reaction residue one-body densities, $\rho^{(j)}_r$, are computed using
spherical Skyrme Hartree-Fock (HF) calculations using the Skyrme SkX interaction
\cite{Bro98}. Such HF calculations have been shown to provide a very good
global description of the root mean squared (rms) sizes \cite{Typ} and radial
forms of the matter and charge distributions of both stable and neutron-proton
asymmetric nuclei \cite{Ric}.

\subsection{Removed-nucleon radial overlaps}
In common with reactions on a composite target, the geometries of the neutron
bound-states potentials, that generate the normalized single-particle overlaps
of the removed nucleons in the projectile ground state, were constrained by
Skyrme (SkX interaction) HF calculations. As discussed in some detail in Ref.
\cite{gade08}, for consistency with the range of the residue-target optical
potential, determined by the $\rho^{(j)}_r$ of the residue, the bound-states
potential geometry is adjusted to reproduce the separation energy and the rms
radius of each single-particle orbital as obtained using the HF. We generate
Woods-Saxon binding potential geometries with reduced radius and diffuseness
parameters ($r_0,a_0) = (r_0,0.7)$ fm. A Thomas-form spin-orbit potential with
a depth of 6 MeV and the same geometry parameters is also included. For neutron
removal from $^{29}$Ne, the so-constrained reduced radii $r_0$ (for further
details see Section III of Ref. \cite{gade08}) are 1.250 fm ($\nu 2s_{1/2}$,
$\nu 2p_{3/2}$, $\nu 1f_{7/2}$) and 1.243 fm ($\nu 1d_{3/2}$, $\pi 1d_{5/2}$),
as were used in Ref. \cite{Kob29}, allowing detailed comparisons with the
results of that analysis. Having determined these HF-constrained geometries,
the bound-state form factors are computed using the empirical ground-state to
ground-state separation energy and the excited shell-model final-state energies.

\subsection{Shell-model calculations}
There are a number of large-basis shell-model calculations, using
effective interactions developed for $sdpf$-shell calculations in the
mass/charge region of interest, e.g. (i) {\sc sdpf-m} \cite{Uts01}, (ii)
{\sc sdpf-u-mix} \cite{Pov12,Cau14} and (iii) {\sc sdpf-ekk} \cite{ekk}.
In our methodology, the projectile and residual nucleus shell-model wave functions
provide: (a) the spectra of the bound final states of the residues, and
(b) from their overlaps, the spectroscopic factors $C^2S(\alpha,n\ell_j)$
between the projectile ground state and all bound residue final states $\alpha\equiv
[E_\alpha^*,J_\alpha^{\pi}]$ that can be reached by direct removal of
a single-neutron with quantum numbers $n\ell_j$. Here, we
will use only the results of the {\sc sdpf-m} interaction \cite{Uts01},
that were used in the carbon and lead target analysis of Ref. \cite{Kob29},
and with which we can thus make detailed comparisons. These shell-model
calculations included the full $sd$-shell plus $1f_{7/2},2p_{3/2}$ states
for neutrons while the protons are confined to the $sd$-shell. Further
comparisons, using the predictions of the alternative shell-model interactions,
will be profitable as experimental data becomes available.

\subsection{Cross-section calculations}
Given the model framework and the inputs discussed above, single-particle
removal cross sections, $\sigma_\text{sp}$ -- the cross sections computed
assuming a normalized nucleon overlap function -- can be computed to each
bound shell-model final state. As clarified earlier,
with a proton target this is the elastic breakup (or diffraction
dissociation) component $\sigma_\text{elas}$, as given by Eq. (6) of Ref.
\cite{Tos01}, there denoted $\sigma_\text{sp}$({\em diff}\,).

The contribution to the partial cross section for a given final
state $\alpha$, of excitation energy $E_\alpha^*$ and spin-parity $J_\alpha^{
\pi}$, due to removal of a nucleon with single-particle quantum numbers
$n\ell_j$, is \cite{HaT03}
\begin{equation}
\sigma_{-1{\rm n}}^{\rm th}(n\ell_j,S_\alpha^*)=\left(\frac{A}{A-1}
\right)^{\!\!{\cal N}}C^2S(\alpha,n\ell_j) \,\sigma_\text{sp}(n\ell_j,S_\alpha^*)
\label{eq:xsec}
\end{equation}
where $\sigma_\text{sp}(n\ell_j,S_\alpha^*)$ is the single-particle cross section.
Here, $S_\alpha^*(=S_{n}+E_\alpha^*)$ is the effective neutron separation
energy to the final state $\alpha$ with $S_{n}$ the empirical ground-state
to ground-state separation energy. ${\cal N}$, in the $A$-dependent
center-of-mass correction to the shell model spectroscopic factor
$C^2S(\alpha,n\ell_j)$, is the number of oscillator quanta associated
with the major shell from which the neutron is removed \cite{Die74}.
Since the odd-$A$, $^{29}$Ne initial state has non-zero spin, the total
theoretical partial cross section to final state $\alpha$, $\sigma_{
-1{\rm n}}^{\rm th} (\alpha)$, is the sum
\begin{equation}
\sigma_{-1{\rm n}}^{\rm th}(\alpha)=\sum \sigma_{-1{\rm n}}^{\rm th}(n
\ell_j,S_\alpha^*) \label{eq:xsecsum}
\end{equation}
of the $\sigma_{-1{\rm n}}^{\rm th}(n\ell_j,S_\alpha^*)$, of Eq.
(\ref{eq:xsec}), from all orbitals $n\ell_j$ with a non-vanishing
shell-model spectroscopic factor to that final state.

\section{Results \label{29sect}}
The earlier carbon target removal reaction analysis of Ref.\ \cite{Kob29}
showed conclusively that the $^{29}$Ne ground-state is a deformed, weakly-bound
$p$-wave neutron-halo system -- also exhibiting a large Coulomb dissociation
cross section on a lead target. The spin-parity of the $^{29}$Ne ground state was
unambiguously determined to be $3/2^-$ using the neutron removal cross sections
on both targets and the momentum distribution of $^{28}$Ne for the carbon-target
induced removal reaction. In particular, the narrow momentum distribution is
typical of a $p$-wave neutron and excluded the earlier ($3/2^+$) evaluated
assignment \cite{ensdf}. The $3/2^-$ assignment is also in line with the enhanced
reaction cross section of $^{29}$Ne \cite{Tak12} that favors a low-$\ell$
valence orbital.

The bound {\sc sdpf-m} $^{28}$Ne($J_\alpha^\pi$) shell-model final states,
their $C^2S(\alpha,n\ell_j)$, and the calculated partial cross sections to these
states from the $^{29}$Ne($3/2^-$) projectile are presented in Table \ref{ne29_plus}.
The results from the analysis of the same transitions on a carbon target can be
found in Table VI of Ref.\ \cite{Kob29} where these shell-model calculations
provided an excellent description of the ground-state and excited-states
inclusive data. Given these earlier data
and its detailed analysis, the present $p(^{29}$Ne,$^{28}$Ne)$pn$ reaction
calculations offer a test case to compare the sensitivities of the carbon target
calculations with those on a proton. To be fully consistent in these comparisons
we take the $^{29}$Ne neutron separation energy to be 963 keV \cite{ame2012}
as used previously.

\begin{widetext}

\begin{table}[!htb]
\caption{Calculated neutron-removal reaction partial and inclusive cross sections
from the 3/2$^-$ {\sc sdpf-m} shell-model state of $^{29}$Ne for a proton target
at a beam energy 240 MeV/nucleon. Tabulated are the neutron-removal partial
cross sections to all particle-bound shell-model $^{28}$Ne final states - states
up to and within the stated error on the first neutron threshold of 3.822(155)
MeV \cite{Wang}. The theoretical cross sections, $\sigma_{-1{\rm n}}^{\rm th}
(n\ell_j,S_\alpha^*)$ and $\sigma_{-1{\rm n}}^{\rm th}(\alpha)$ include the
center-of-mass correction factor $\left[A/(A-1\right)]^{\cal N}$ to the
shell-model spectroscopic factors, as per Eq. (\ref{eq:xsec}).
\label{ne29_plus} }
\begin{ruledtabular}
\begin{tabular}{lccccccc}
Reaction & $E^*_{\alpha}$ & $J_\alpha^{\pi}$ & $n\ell_j$ & $\sigma_{\rm sp}
(n\ell_j,S_\alpha^*)$ & $C^2S(\alpha,n\ell_j)$ & $\sigma_{-1{\rm n}}^{\rm th}
(n\ell_j,S_\alpha^*)$& $
\sigma_{-1{\rm n}}^{\rm th}(\alpha)$  \\
 & (MeV) &  &  & (mb) &  & (mb) & (mb)  \\ \hline
C[$^{29}$Ne(3/2$^-$),$^{28}$Ne($\alpha$)]
& 0.00 & 0$^+_{1}$ & $2p_{3/2}$ & 21.19 & 0.438 & 10.31 &10.31   \\
$S_{n}$($^{29}$Ne) = 0.963 MeV
& 1.36 & 2$^+_{1}$ & $2p_{3/2}$ & 18.51 & 0.072 &  1.48  &4.02    \\ 		
&      &           & $1f_{7/2}$ & 13.69 & 0.167 &  2.54 &        \\ 		
& 2.21 & 0$^+_{2}$ & $2p_{3/2}$ & 17.43 & 0.005 &  0.10 & 0.10    \\ 		
& 2.76 & 4$^+_{1}$ & $1f_{7/2}$ & 12.97 & 0.417 &  6.01 & 6.01    \\ 		
& 2.99 & 2$^+_{2}$ & $2p_{3/2}$ & 16.63 & 0.066 &  1.22 & 1.43    \\ 		
&      &           & $1f_{7/2}$ & 12.87 & 0.015 &  0.21 &         \\ 		
& 3.57 & 2$^-_{1}$ & $2s_{1/2}$ & 15.04 & 0.036 &  0.58 & 1.01    \\ 		
&      &           & $1d_{3/2}$ & 11.36 & 0.035 &  0.43 &         \\ 		
& 3.69 & 3$^-_{1}$ & $1d_{3/2}$ & 11.30 & 0.236 &  2.86 & 3.09    \\ 		
&      &           & $1d_{5/2}$ & 12.37 & 0.017 &  0.23 &         \\ 		
& 3.98 & 2$^+_{3}$ & $2p_{3/2}$ & 15.81 & 0.153 &  2.69 & 2.76    \\ 		
&      &           & $1f_{7/2}$ & 12.46 & 0.005 &  0.07 &         \\ 		
& 3.99 & 4$^+_{2}$ & $1f_{7/2}$ & 12.46 & 0.258 &  3.57 & 3.57
\smallskip  \\ \hline
&      & \multicolumn{3}{c}{Excited states sum}   &   &  & 22.0\\
\hline		
&      & \multicolumn{3}{c}{Inclusive}   &    &          & 32.3 \\
\end{tabular}
\end{ruledtabular}
\end{table}

\end{widetext}

We see that the calculated inclusive cross section is 32.3 mb and its
ground-state component is 10.3 mb, to be compared with those for the
carbon target, of 69.0 mb, and 31.6 mb. In considering these absolute
cross-sections and their likely uncertainties with regard the model
inputs, we note the most recent evaluated $^{29}$Ne ground-state separation
energy is $S_n= 971(196)$ keV \cite{Wang}. Repeating the present calculations
for this value of $S_n$, and its quoted uncertainty, the calculated
ground-state and excited-states-inclusive cross sections are $10.30^{+0.29
}_{-0.25}$ mb and $22.0^{+0.19}_{-0.18}$ mb, respectively, and remain
in agreement with the values shown in Table \protect\ref{ne29_plus}. So, the
$\approx 20\%$ uncertainty on the ground-state separation energy results
in cross-section changes smaller than the errors on typical measurements
by about a factor of three.

\subsection{Neutron orbital sensitivity}
It is widely expected, see e.g. Ref. \cite{qe1} and references therein,
that the use of a proton target will increase the sensitivity of the reaction
to the overlap functions (removed-nucleon wave functions) in the projectile
interior. The extent of this sensitivity will depend on the beam energy and
reflects a modest increase in the transparency and penetration of the proton
at the projectile surface. The reaction on the proton target is thus
less surface localised than when very highly-absorptive optical interactions
with a nuclear target are present. We quantify this effect in the case of
weakly-bound, neutron-rich $^{29}$Ne.

The partial cross section results of Table \ref{ne29_plus} and those of Table
VI of Ref.\ \cite{Kob29}, that involve several $\ell$ values and a range of
separation energies, are compared. One clear result from this comparison is
that the proton target reaction is significantly less effective in overlapping
with and removing neutrons from the more spatially extended configurations, i.e.
the more weakly-bound and low-$\ell$ orbitals, such as the ground-state to
ground-state removal. To elucidate these different sensitivities Table
\ref{tbl:percent} summarizes the two cases and separates the calculated
contributions to the bound-states-inclusive cross sections (on the proton
and carbon targets) according to their $\ell$-value. The table also presents
the percentage contributions to the inclusive cross section for removal from
the lower $\ell$, $s+p$-state, and higher $\ell$, $d+f$-state, single-particle
orbitals; showing the relatively higher percentage of cross section, 49\%,
due to the $d$- and $f$-wave orbitals in the proton target case -- as compared
to 35\% for the carbon target. The ratios of the cross sections from each
$\ell$ on the proton and carbon targets are also shown. The smallest p:C
ratio (0.36), for $p$-wave contributions, is dominated by the weakly-bound
ground-state to ground-state $2p_{3/2}$ neutron component. This $\ell=1$
ground-state transition alone has ratio 10.3 : 31.6 = 0.33, showing the
greater efficacy of the carbon target in removing such halo-like nucleon
configurations. The ratios for $\ell=2$ and $\ell=3$ orbitals is 0.65, a
factor of two difference. Confirmation of such sensitivities is possible
given current experimental capabilities.

\begin{table}[!htbp]
\begin{ruledtabular}
\begin{center}
\caption{Calculated neutron-removal cross sections from $\prj$. The
$\sigma_{-1{\rm n}}^{\rm th}(\ell)$, for $s,p,d$ and $f$-wave orbitals,
present the contributions to the inclusive cross-section arising
from each $\ell$-value of the removed nucleon's orbital, on the
proton (from Table \protect\ref{ne29_plus}) and carbon targets
(from Table VI of Ref. \cite{Kob29}). The ratios of these cross
sections on the two targets and the percentage contributions
to the inclusive cross section to bound $^{28}$Ne shell model final
states from low-$\ell$ ($2s_{1/2}$ and $2p_{3/2}$) and higher-$\ell$
($1d_{3/2}$, $1d_{5/2}$ and $1f_{7/2}$) neutron orbitals are also
shown.}
\begin{tabular}{cccc}
    & p target & C target &Ratio(p:C)  \\\hline
$\sigma_{-1{\rm n}}^{\rm th}(s)$ (mb)&  0.58    &  1.17&0.50   \\	
$\sigma_{-1{\rm n}}^{\rm th}(p)$ (mb)& 15.80    & 43.36&0.36   \\	
$\sigma_{-1{\rm n}}^{\rm th}(d)$ (mb)&  3.51    &  5.38&0.65   \\	
$\sigma_{-1{\rm n}}^{\rm th}(f)$ (mb)& 12.40    & 19.10&0.65   \\	
Inclusive (mb)& 32.3    & 69.0 &  0.47 \\\hline	
$s+p$ : $d+f$ (\%)             & 51 : 49  & 65 : 35 \\
\end{tabular}\label{tbl:percent}
\end{center}
\end{ruledtabular}
\end{table}

Referring to Table \ref{ne29_plus}, the calculated percentage of the
inclusive cross section from the ground-state:excited-states transitions
is $32 : 68$\% on the proton target -- whereas on carbon these contributions
are comparable, with 46 : 54\%. Again, these differences are significant
and accessible to experimentation. The ratio from the carbon target data
was 49(9) : 51\%, consistent with the model calculations. Again, the larger
proton target cross-section contribution from excited states confirms
the expected increased sensitivity to the more spatially localized,
higher-$\ell$ and more well-bound orbital components in the $^{29}$Ne
ground-state.

\subsection{Parameter sensitivity: NN S-matrix \label{sens}}
As part of this benchmark study, it is also useful to clarify the sensitivity
of the calculated removal reaction single-particle cross sections, $\sigma_
\text{sp}$, to the parameters used in the description of the NN profile
function and S-matrix; namely the $\alpha_{jp}$ and the associated $\beta_{jp}$.
This assessment is presented in Table \ref{nnsens}, which shows the
$\sigma_\text{sp}$ calculated for a selection of the important {\sc sdpf-m}
$^{28}$Ne shell-model final states from Table \protect\ref{ne29_plus}. The
calculations use: (i) the $\alpha_{np}$ values from the Ray
tabulation \cite{Ray}, as used above, and (ii) the extreme choice that
$\alpha_{np}=0$. In both cases the $\beta_{np}$ were determined
using Eq. (\ref{betaeq}). For the Ray value, $\alpha_{np}=0.533$, $\beta_{np}
=0.0949$ fm$^2$, which corresponds to a Gaussian range, $\gamma_{np}=
\sqrt{2 \beta_{np}} =0.436$ fm. Taking $\alpha_{np}=0$, then $\beta_{np}=
0.0739$ fm$^2$, or $\gamma_{np}=0.384$ fm.

\begin{table}[!htbp]
\caption{Calculated single-particle cross sections for selected {\sc sdpf-m}
shell-model $^{28}$Ne final states at 240 MeV/nucleon, when using different parameters
$\alpha_{np}$, and their associated $\beta_{np}$, computed via Eq. (\ref{betaeq}),
in the NN S-matrix. The $\alpha_{np}=0.533$ cross sections are as in
Table \ref{ne29_plus}. All calculations use $\sigma_{np} =37.14$ mb.
 \label{nnsens}}
\begin{ruledtabular}
\begin{center}
\begin{tabular}{llccc}
State& $E^*$ (MeV)&$n\ell_j$& $\sigma_\text{sp}$ (mb) & $\sigma_
\text{sp}$ (mb) \\
&&$ $&$\alpha_{np}=0.533$ & $\alpha_{np}=0.0$ \\
\hline
$0^+$  & 0.00&$2p_{3/2}$& 21.19 & 20.67 \\
$2^+$  & 1.36&$2p_{3/2}$& 18.51 & 18.28 \\
       &     &$1f_{7/2}$& 13.69 & 13.76 \\
$4^+$  & 2.76&$1f_{7/2}$& 12.97 & 13.07 \\
$3^-$  & 3.69&$1d_{3/2}$& 11.30 & 11.38 \\
       &     &$1d_{5/2}$& 12.37 & 12.44 \\
\end{tabular}
\end{center}
\end{ruledtabular}
\end{table}

Inspection of these $\sigma_\text{sp}$ shows that the proton target
calculations are extremely insensitive to details of the $\alpha_{jp}$
(and $\beta_{jp}$) parameters in the NN S-matrix parameterization, i.e.
to details of the NN system beyond the experimentally-constrained NN
total cross section $\sigma_{np}$. This provides confidence in the
stability of the calculated values with respect to realistic variations
of these model inputs.

\section{Summary Comments}
Intermediate-energy nucleon removal reactions are used as a valuable
and efficient tool to study the single-particle degrees of freedom and
their evolution with mass and charge in nuclei far from stability. We
have presented model calculations for such reactions on a proton target,
for which a basic input is the NN collision S-matrix. We show that, in the
model used, calculations are highly insensitive to details of the parameters
used to describe the NN profile function, beyond the empirically
well-determined NN total cross sections. We compare the computed cross
sections to the {\sc sdpf-m} shell-model $^{28}$Ne final states in the
$p(^{29}$Ne,$^{28}$Ne)$pn$ reaction at 240 MeV/nucleon with the earlier
analysis of the reaction on a carbon target at the same incident energy
per nucleon.

As anticipated, the calculated final-state cross sections on the proton
target show a significantly different sensitivity to the $\ell$-value
and separation energy of the orbital of the removed nucleon -- with
relatively enhanced cross sections for larger-$\ell$ and more bound
orbitals. The proton target efficacy is less for removals from low-$
\ell$ and more halo-like valence orbitals -- compared to reactions on
a carbon target. These differences are quantified and predicted to
be significant. Measurements of these differences in sensitivity, e.g.
of the ratios of measured final-state-exclusive cross sections on proton
and carbon targets, could help to experimentally identify states having
a halo-like character. Future analyses, where measured and calculated
partial or inclusive cross sections can be compared, can
test these quantitative model predictions. Such data should include
both neutron and proton removals from projectile nuclei with an extended
range of $(N,Z)$ values. These will include cases of strongly-bound
nucleons and so can probe the reaction model predictions over a significant
range of nucleon separation energies.

\begin{acknowledgments}
Valuable discussions with Professor Takashi Nakamura and members of his
laboratory, including comments on an earlier draft of this work are
acknowledged. Visiting researcher support at Tokyo Institute of Technology,
from MEXT  KAKANHI Grant No. JP18H05400 and the support of the Science and
Technology Facilities  Council (U.K.) Grant No. ST/L005743/1 are acknowledged.
\end{acknowledgments}

\end{document}